\documentclass[12pt]{article}
\usepackage{amsmath}
\usepackage{multicol, multirow,comment}
\usepackage{graphicx,psfrag,epsf}
\usepackage{enumerate}
\usepackage{array}
\usepackage[numbers]{natbib}
\usepackage{url} 
\usepackage{longtable}
\usepackage{booktabs}
\usepackage{subcaption}
\usepackage{chngcntr}
\usepackage{lineno}
\newcommand{\blind}{1}

\begin{document}

\def\spacingset#1{\renewcommand{\baselinestretch}%
{#1}\small\normalsize} \spacingset{1}

\if1\blind
{

  \title{\bf Comparing Step Counting Algorithms for High-Resolution Wrist Accelerometry Data in NHANES 2011-2014}
  \author{Lily Koffman$^{1*}$\\
         Ciprian Crainiceanu$^1$\\
         John Muschelli$^1$\\
         \textbf{Running title}: Open-Source Step Counts in NHANES\\
         $^1$Department of Biostatistics, Johns Hopkins Bloomberg School of Public Health\\
         $^*$Corresponding author: lkoffma2@jh.edu, 615 N Wolfe St, Baltimore MD 21231 }
  \maketitle
} \fi

\if0\blind
{
  \bigskip
  \bigskip
  \bigskip
  \begin{center}
    {\LARGE\bf Heterogeneity of Open-Source Step Counting Algorithms in the National Health and Nutrition Examination Survey}
\end{center}
  \medskip
} \fi

\clearpage 

\begin{abstract}
\noindent \textbf{Purpose:}\\ To quantify the relative performance of step counting algorithms in studies that collect free-living high-resolution wrist accelerometry data and to highlight the implications of using these algorithms in translational research.  \\\\
\textbf{Methods:}\\ Five step counting algorithms (four open source and one proprietary) were applied to the publicly available, free-living, high-resolution wrist accelerometry data collected by the National Health and Nutrition Examination Survey (NHANES) in 2011-2014. The mean daily total step counts were compared in terms of correlation, predictive performance, and estimated hazard ratios of mortality.  \\\\
\textbf{Results:} \\ The estimated number of steps were highly correlated (median=$0.91$, range $0.77$ to $0.98$), had high and comparable predictive performance of mortality (median concordance$=0.72$, range $0.70$ to $0.73$). The distributions of the number of steps in the population varied widely (mean step counts range from $2{,}453$ to $12{,}169$). Hazard ratios of mortality associated with a $500$-step increase per day varied among step counting algorithms between HR=$0.88$ and $0.96$, corresponding to a 300\% difference in mortality risk reduction ($[1-0.88]/[1-0.96]=3)$.\\\\
\textbf{Conclusion:}\\ Different step counting algorithms provide correlated step estimates and have similar predictive performance that is better than traditional predictors of mortality. However, they provide widely different distributions of step counts and estimated reductions in mortality risk for a $500$-step increase.

\end{abstract}

\noindent%
\vfill

\newpage
\spacingset{1.45} 
\section{Introduction}
\label{sec:intro}
Objective estimation of the number of steps using high-resolution wrist accelerometry data has become increasingly important because: (1) large studies, such the National Health and Nutrition Examination Survey (NHANES) and UK Biobank now routinely collect, store, and share publicly high-resolution accelerometry data from wrist-worn wearable devices on tens or hundreds of thousands of study participants for many days at a time; (2) these data are often linked to high quality demographic, behavioral, and health information; (3) the number of daily steps is easy to understand and communicate and can be used as a target for interventions; (4) self-reported walking time or number of steps is subjective and often affected by uncontrollable bias and measurement error \cite{Prince2008}; and (5) accelerometry data and its summaries have been shown to be highly predictive of current and future health status \cite{leroux2021quantifying, ac_frag, Smirnova2019}.

Wrist-worn accelerometers provide measurements of the acceleration of the device attached to the wrist of a person. Specifically, data generated by wrist-worn accelerometers represent the acceleration expressed in Earth's gravitational units ($g=9.81$ $m/s^2$) along three axes at high resolution, typically between $20$ and $100$ observations per second ($20$ to $100$Hz). The three axes represent the frame of reference of the device, which is related to, affected by, but not the same as a particular orientation of the hand and/or wrist. Step counting algorithms are applied to these high resolution accelerometry data to produce an estimated number of steps at the second, minute and/or day level. The downside is that these algorithms are often developed and validated on small data sets, in well controlled environments (e.g., walking on a treadmill) and with a limited number of tasks that do not represent the activity heterogeneity in the free-living environment \cite{steps_koffman}. The upside is that some of these algorithms are open source, which could improve the harmonization of the ``step count" across studies. Ultimately, this could lead to clear, evidence-based age- and sex-specific step counts in the population as well as estimation of the reduction in hazard associated with a specific increase in the number of steps per day (e.g. $500$). However, the output of these algorithms are estimators, not true ``step counts", and they vary depending  on the particular type and version of the algorithm used.

In this paper we compare the results of different step counting algorithms deployed on the same high-resolution acceleromery data collected in the NHANES 2011-2014 study and released in December 2022. The NHANES sample is representative of the US population and the devices were worn continuously for up to seven full days in the free-living environment. The size of the compressed data set is $20$ terabytes and contains information for $14{,}693$ study participants. Deploying the algorithms on this data took the equivalent of approximately 2.5 years of computation time, with some algorithms being much faster than others (see Supplementary Table~\ref{tab:compute}).  Because the gold standard for the number of steps is not available in this data, we focus on studying the  distributions (in the population and by age), correlation, predictive performance, and the estimated hazard ratios of mortality for an increase of $500$ steps per day.

Understanding these differences in these quantities is crucial, as substantial confusion persists in the literature about the population and age/sex-specific reference distributions of the number of steps as well as the health effects associated with a specific increase in the number of steps. A big contributor to this confusion could be the bias induced by the step counting algorithms, even though the targets of inference (estimands) are thought to be well defined and intuitive. We summarize relevant estimates of step counts from the published literature in Table~\ref{tab:methods}. Saint-Maurice \cite{saint-maurice_association_2020} reported an average of $9{,}124$  steps in NHANES 2003-2006 using hip-worn accelerometers among $4{,}840$ US adults over the age of $40$. Tudor-Locke et al. \cite{tudor-locke_accelerometer-determined_2009} used only NHANES 2005-2006 and reported an average number $9{,}676$ steps among $3{,}744$ US adults over the age of $20$. Interestingly, Tudor Locke et al. \cite{tudor-locke_accelerometer-determined_2009} also reported an average of $6{,}540$ steps ($32$\% fewer) by slightly tweaking the step counting algorithm. Small et al. \cite{stepcount} estimated an average of $9{,}352$  steps in the UK Biobank using a wrist-worn accelerometer and open source software on $75{,}943$ UK adults over the age of $40$. This matches the average number of steps reported by Tudor-Locke et al. \cite{tudor-locke_accelerometer-determined_2009} in a comparable US population. Using the same population and data as Small et al. \cite{stepcount}, Chan et al. \cite{watchwalk} reported an average of $8{,}016$ steps, or $14$\% fewer. Lee et al. \cite{lee_association_2019} reported an average of $5{,}499$ number of steps in the Women's Health Study using hip-worn accelerometers among $16{,}741$ US women over the age of $62$. This is about $40$\% fewer steps than reported by Saint-Maurice et al. \cite{saint-maurice_association_2020} in a younger population. Master et al. \cite{master_association_2022} reported an average of $7{,}731$ steps in the All of Us study using wrist-worn Fitbits among $6{,}042$ US adults with a median age of $57$. This is $15$\% fewer steps than reported by Saint-Maurice et al., \cite{saint-maurice_association_2020} in a US population comparable in terms of age.

These differences may be due to the variations in the population, device sensitivity, device location, or algorithmic bias. Regardless of the cause, the differences have consequential implications. For example, the close consensus around an average of $9{,}200$ steps per day between NHANES 2003-2006 and UK Biobank would correspond to an average of $4.0$ to $4.5$ miles ($6.5$ to $7.2$ kilometers), or $1.2$ to $1.5$ hours of walking per day. {\it This would be excellent news about the physical activity (PA) in the US and the UK, as the current Physical Activity Guidelines for Americans recommend 2.5 to 5 hours of moderate intensity activity per \bf{week}} and estimate that only 20\% of Americans are meeting these recommendations \cite{pa_guidelines_2018}. Brisk walking (2.5-4.0 miles per hour) is a moderate intensity activity; we will not discuss walking speed in this paper. However, overestimating the true average has negative consequences; an average of $9{,}200$ steps could seem out of reach and thereby could efforts to increase activity. Moreover, an increase of $500$ steps (about four to seven minutes of walking) from an average of $9{,}200$ steps would correspond to a smaller relative change in behavior than if the true average number of steps were $5{,}000$.

The goals of this paper are to: (1) obtain and publish minute-level step counts from the raw accelerometry in NHANES 2011-2014 using open source step counting methods; (2) describe patterns in step counts by age; (3) investigate the agreement between different step counting methods and their association with other summaries of accelerometry data (namely, Monitor Independent Movement Summary [MIMS] and Activity Counts [AC]); (4) compare the predictive performance of step estimates with other accelerometery-derived and non-accelerometry covariates on mortality; and (5) quantify the association between step counts and mortality in NHANES.

\section{Methods}\label{sec:methods}

\subsection{Study population}\label{subsec:study_population}
NHANES is a nationally representative study of about $5{,}000$ Americans per year. The study includes demographic, socioeconomic, dietary, and health-related questions; data from the study are publicly available. All NHANES study participants age six and older (2011-2012, Wave G) or age three or older (2013-2014, Wave H) were asked to wear an ActiGraph GT3X+ on the non-dominant wrist starting on the day of their exam at the NHANES Mobile Examination Center. They were instructed to wear the device at all times for seven consecutive days and remove the device on the morning of the ninth day. The devices collected triaxial acceleration at $80$ Hz. A total of $6{,}917$ individuals in NHANES 2011-2012 and $7{,}776$ individuals in NHANES 2013-2014 had accelerometer data for analysis. 96\% of participants with data wore the device until the ninth day and approximately 2\% of participants wore the device for fewer than seven days \cite{paxming, paxminh}. More details on the NHANES accelerometer procedures can be found at \url{https://wwwn.cdc.gov/nchs/data/nhanes/2011-2012/manuals/Physical_Activity_Monitor_Manual.pdf}. 

\subsection{Inclusion criteria}\label{subsec:inclusion_criteria}

A machine learning algorithm was used by NHANES to classify each minute of the day as wake wear, sleep wear, nonwear, or unknown \cite{nhanes_wear_algo}. Minutes characterized as unknown were often short in duration (mean of $1.17$ minutes) and the majority of the time were sandwiched between either two periods of wake wear ($46$\% of unknown bouts) or two periods of sleep wear ($28$\% of unknown bouts) (see Supplementary Table~\ref{tab:tpm}). Thus, for the purpose of this paper unknown minutes were counted as wear.  Minute-level data quality flags were also provided based on a number of rules indicating if any issues were detected \cite{nhanes_pam}. A single day was considered valid if it met the following three conditions: (1) at least $1{,}368$ minutes ($95$\% of a full day) were classified as wake wear, sleep wear, or unknown and did not have any data quality flags; (2) at least seven hours ($420$ minutes) of the day were classified as wake wear; and (3) at least seven hours ($420$ minutes) had non-zero activity level data (MIMS, see below). Data from an individual were considered valid if they had at least three days with valid wear. Sensitivity analyses demonstrated that results were similar when including individuals with at least one day of valid wear (see Supplementary Table~\ref{tab:sens_univariate} and Supplementary Table~\ref{tab:sens_multivar}).

\subsection{Accelerometery-derived physical activity summaries}\label{subsec:summaries}

\subsubsection{MIMS and Activity Counts}\label{subsubsec:MIMS}
Monitor Independent Movement Summary (MIMS), an open source summary of raw accelerometry data \cite{MIMS}, were calculated from the raw accelerometry data and provided by NHANES at the minute level in the \texttt{PAXMIN} files. Log base 10 MIMS, referred to throughout the rest of the manuscript as ``log10 MIMS", were calculated by applying the $\text{log}_{10}(1+\text{MIMS})$ transformation at the minute level. One is added so that zero values are not excluded. 

The ActiGraph Activity Count (AC) has been used in thousands of manuscripts. While previously only available as a proprietary algorithm, it was recently made open-source \cite{activity_counts}. This algorithm was implemented in R \cite{R} using the package \texttt{agcounter} \cite{agcounter}, which wraps ActiGraph's Python code to create activity counts \cite{agcounts}, and applied to the raw accelerometry data to calculate activity counts at the second level.  Log base 10 AC, referred to throughout the rest of the manuscript as ``log10 AC", were calculated by summing the AC at the second level to obtain minute level ACs and applying the $\text{log}_{10}(1+\text{AC})$ transformation at the minute level. 

\subsubsection{Steps}\label{subsubsec:steps}
Four open source step counting methods: ADEPT \cite{adept}, Oak \cite{oak}, Verisense \cite{verisense_2022, verisense_github}, and Stepcount \cite{stepcount}; and one proprietary algorithm, Actilife \cite{Actilife}, were applied to the raw NHANES data to estimate steps for each participant. The R package \texttt{adept} \cite{adept} was used to implement ADEPT, the R package \texttt{walking} \cite{walking_pkg} was used to implement Oak and Verisense, and the R package \texttt{stepcount} \cite{stepcount_pkg} was used to implement both the random forest (RF) and self-supervised learner (SSL) versions of Stepcount \cite{stepcount}. Both the original \cite{verisense_github} and revised \cite{verisense_2022} version of Verisense were used; for more details on each algorithm, see \cite{steps_koffman}. The estimated steps for the entire NHANES data set using these algorithms will be made available on Physionet at the time of publication; the associated software, and an accompanying vignette are publicly available at \url{https://github.com/lilykoff/nhanes_steps_mortality}.

For all step estimates, MIMS, log10 MIMS, AC, and log10 AC, values were first summed at the minute level. To obtain day totals, values from valid minutes were summed over each day; again, valid minutes were defined as minutes that did not have any data quality flags and were classified by the wear algorithm as wake wear, sleep wear, or unknown. Day totals were then averaged across valid days to obtain one summary for each individual and physical activity variable.  These represent values for the ``average day''.

\subsection{Mortality and other covariates}

The following variables from the NHANES 2011-2014 data were extracted: age, sex (male/female), race/ethnicity (non-Hispanic white, non-Hispanic Black, Mexican-American, Other Hispanic, Other (including multi-race)), education level (less than high school, high school/high school equivalent, more than high school), body mass index (BMI) category (underweight, normal, overweight, obese), diabetes, coronary heart disease (CHD), congestive heart failure (CHF), heart attack, stroke, cancer, alcohol consumption (never drinker, former drinker, moderate drinker, heavy drinker, missing alcohol), cigarette smoking (never smoker, former smoker, current smoker), mobility problem, and self-reported health status (poor, fair, good, very good, excellent). We refer to these variables throughout the manuscript as ``traditional predictors.''

Mortality data was obtained from the national death registries public-use linked mortality files and is available publicly for participants $\geq$ 18 years old (see \url{https://www.cdc.gov/nchs/data-linkage/mortality-public.htm}). The mortality files include vital status (assumed alive or assumed deceased) and person-months of follow-up time from the mobile examination center visit to the date of death or the end of the mortality follow-up period (December 31, 2019). Mortality data were merged with the covariates data. 

\subsection{Statistical analysis}

For each step counting algorithm, the survey weighted mean and standard error of step counts by age was estimated using the \texttt{survey} \cite{survey_pkg} package in R following the NHANES weighting guidelines. The means and associated $95$\% confidence intervals (CI) were smoothed as functions of age using locally weighted smoothing \cite{loess}. The agreement between different step counting methods was quantified using the Spearman and Perason correlations between the mean daily step counts from each method, mean daily AC, and mean daily MIMS.

To evaluate the predictive performance of all covariates we first fit separate univariate Cox proportional hazards regression models on  mortality for each predictor among individuals between $50$ and $79$ years old at screening (n = 3638; number of deaths = 412; median follow up time = 6.75 years). Age was restricted to a maximum of 79 because individuals 80 and over are topcoded at 80 years of age in NHANES. A sensitivity analysis was performed including individuals 80 and older and results were similar (see Supplementary Table~\ref{tab:sens_univariate} and Supplementary Table~\ref{tab:sens_multivar}). For each model, a ten-fold survey weighted cross-validated concordance (cvC) \cite{concordance} was calculated.  This calculation was repeated 100 times and the average of these 100 cvC values is reported for stability \cite{ms_ukb, pd_ukb, repeat_cv}. Since all PA covariates were right skewed, they were Winsorized at the 99th percentile before model fitting (see Supplementary Figure~\ref{fig:distributions}) 

A series of four models was used to investigate the association between covariates and mortality. The first model included traditional predictors. The second model contained the same covariates as the first model plus the mean total daily MIMS. The third model contained the same covariates as the first model and mean total steps, as estimated by the step algorithm with the highest univariate cvC. The final model contained the same covariates as the first model, total MIMS, and total steps. The 100-times repeated 10-fold cvC is calculated for each model, along with the coefficient estimate and p-value associated with the steps variable. 

To estimate the association of step counts with mortality, separate multivariable Cox proportional hazards regression models were fit for each step count algorithm. All models included the traditional predictors and one step count algorithm. Separate models were fit using both raw and standardized (centered and scaled) step counts. The covariate-adjusted hazard ratio of mortality was calculated for: (1) an increase of $500$ in mean daily steps; and (2) a one standard deviation increase in mean daily steps. 

All statistical analyses were performed using R version 4.4.1. The R packages \texttt{tidymodels} \cite{tidymodels_pkg} and  \texttt{survival} \cite{survival-package} were used for model fitting and cross-validation. 

\section{Results}\label{sec:results}
 After applying the inclusion criteria described in Section~\ref{subsec:inclusion_criteria} and excluding everyone under the age of $18$, the analytic sample contained $4{,}303$ individuals from 2011-2012 and $4{,}361$ from 2013-2014. For the mortality prediction analysis, individuals who were younger than 50, older than 79, or had missing covariates were excluded, resulting in an analytic sample of $3{,}368$ ($1{,}795$ from 2011-2012 and $1{,}843$ from 2013-2014). Table~\ref{tab:pop_chars} presents survey weighted characteristics of individuals 18 years and older with valid accelerometry data by wave and overall. Supplementary Figure~\ref{fig:supp-consort} summarizes the inclusion/exclusion process, and Supplementary Table~\ref{tab:mortality_tab1}
 presents the non-survey weighted characteristics of the individuals included in the mortality analysis.

\subsection{Step counts by age}\label{subsec:stepbyage}
Table~\ref{tab:steps_pa} presents the estimated survey weighted means and standard deviations for step counts from each algorithm by wave and age category. Across all groups, Actilife consistently estimates the highest mean step count, although the estimates from Actilife, Oak, and Stepcount RF are similar. ADEPT estimates the lowest step counts. For all methods, the average number of steps decreases for individuals aged 50 years and older. Verisense (original and revised) and stepcount SSL produce similar step count estimates. The final column of Table~\ref{tab:steps_pa} presents the absolute difference divided by the mean in the 2011-2012 and 2013-2014; a higher number indicates larger differences between waves. Interestingly, the percent difference in step estimates between waves among all adults ranges from 3.1\% (Actilife) to 8.5\% (ADEPT), and differences observed for any step algorithm are larger than those observed for MIMS, AC, log10 MIMS, or log10 AC (range 0.84 to 1.8\%). 

\begin{figure}[!htbp]
    \centering
    \includegraphics[width=.95\textwidth]{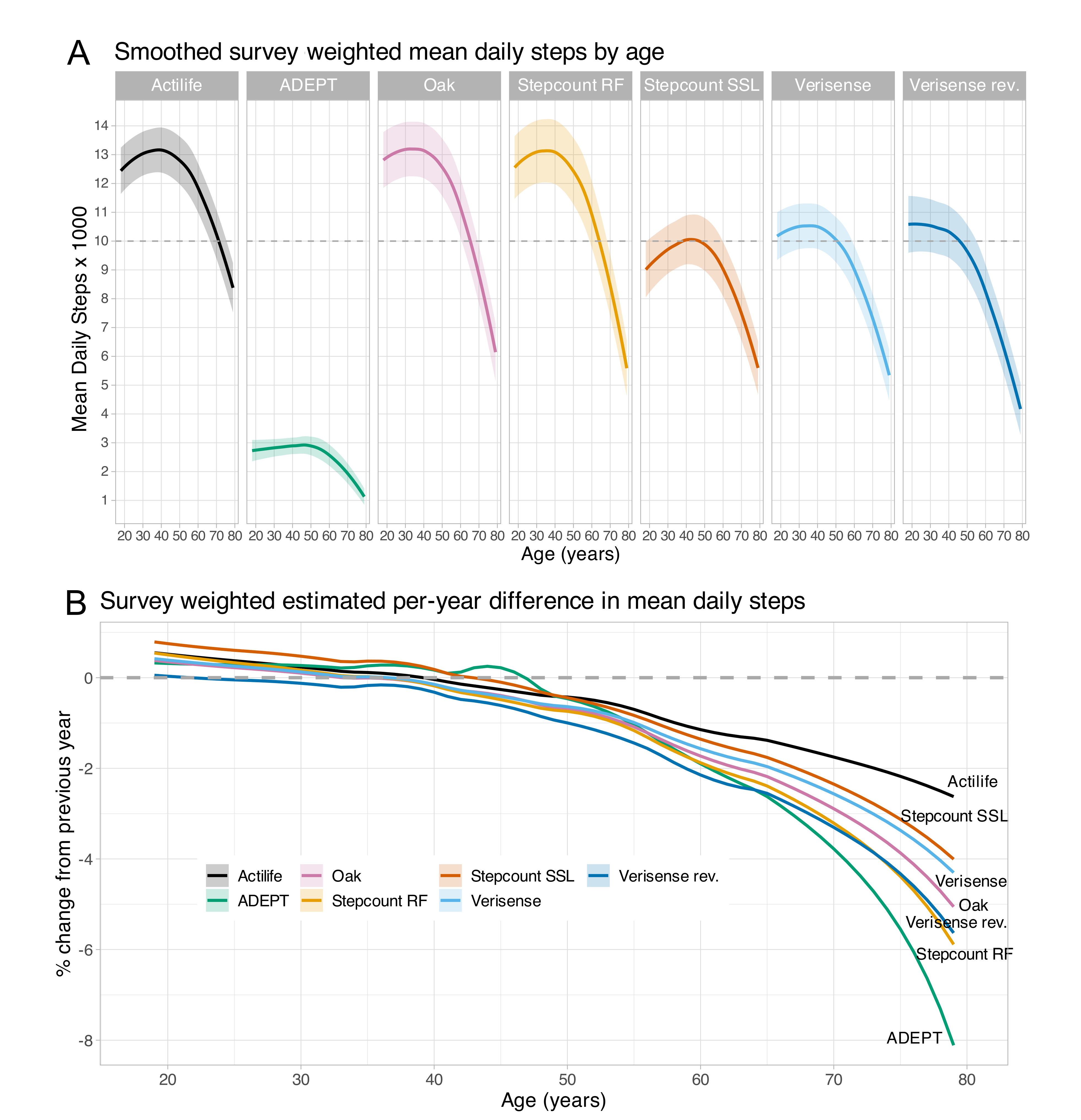}
    \caption{Panel \textbf{A}: survey weighted mean and associated 95\% confidence intervals for total daily step counts, by algorithm and age. The horizontal dashed line indicates the 10,000 steps for reference. Panel \textbf{B}: percent change in smoothed survey weighted mean daily steps for each year compared to the previous year, by algorithm. The horizontal dashed line indicates no change (0\%) for reference.}
    \label{fig:mean_pa}
\end{figure}

Figure ~\ref{fig:mean_pa} panel A displays the smoothed, survey weighted mean step counts and associated $95$\% confidence intervals by age for each step counting method. Actilife ($13,000$ at age $40$ and $8,500$ at age $80$), Oak ($13,000$ at age $40$ and $6,000$ at age $80$), and Stepcount RF ($13,000$ at age $40$ and $5,500$ at age $80$) have the highest step count estimates across all ages. These methods estimate a loss of approximately $110$ steps per year from age $40$ to age $80$, or $1$\% per year.  Stepcount SSL and both Verisense versions estimate around $10{,}000$ steps at age 40 and $4{,}000$ to $5{,}500$ steps at age 80; a loss of approximately $110-150$ steps per year. ADEPT estimates the lowest average step count ($3,000$ at age $40$ and $1,000$ at age $80$); a loss of $50$ steps or approximately 1.6\% per year from age $40$ to age $80$.

Panel B plots the estimated per-year percent difference in mean daily steps, which are consistent with $2$\% to $1$\% increases between ages $20$ and $30$, small changes between $30$ and $40$, $1$ to $4$\% decreases between $50$ and $70$, and $2$ to $4$\% decreases between $70$ and $75$. Steeper declines occur after age $75$, especially for ADEPT.

\subsection{Correlation between step counting algorithms}\label{subsec:correlation}

\begin{figure}[!htbp]
    \centering
    \includegraphics[width=.9\textwidth]{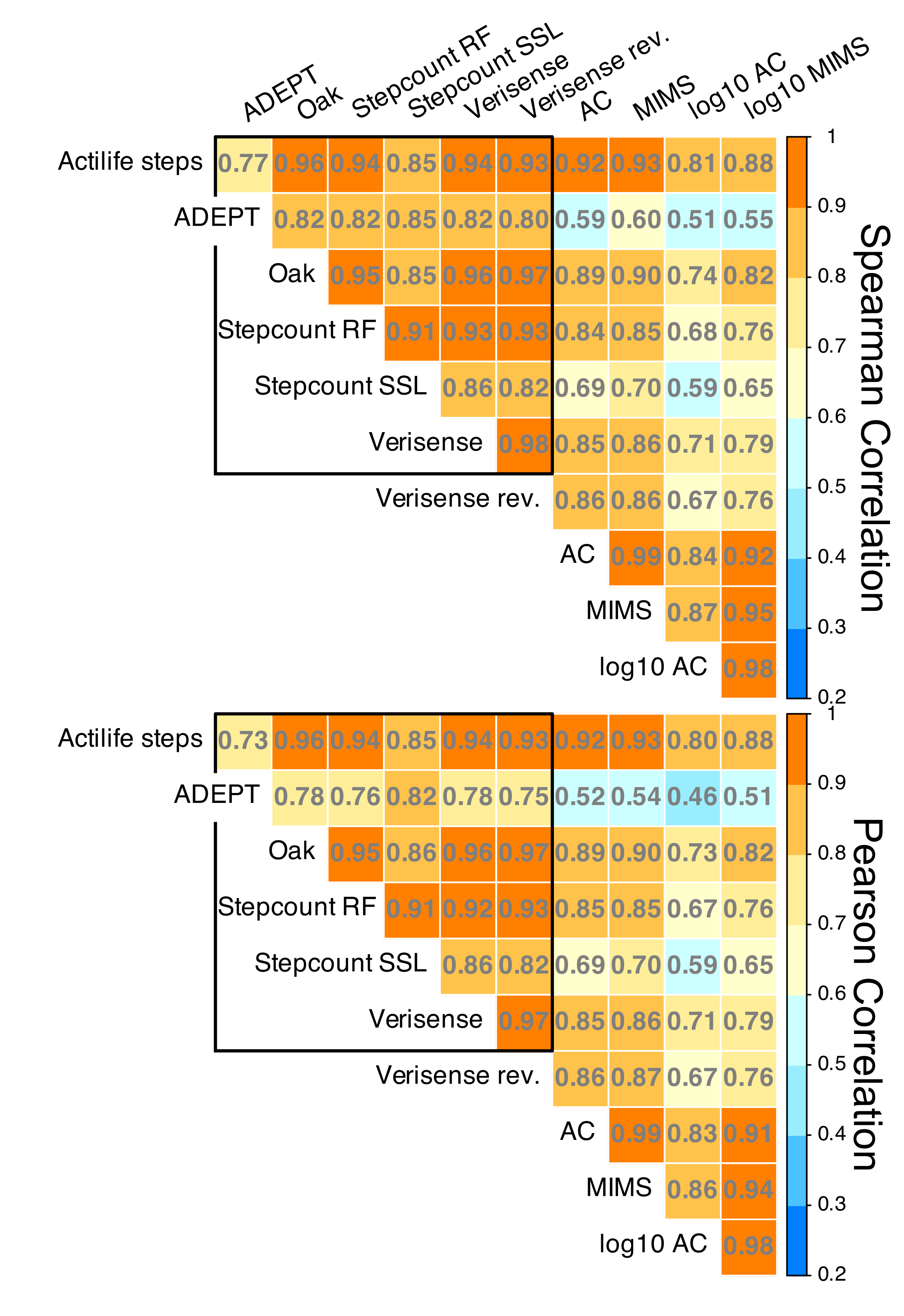}
    \caption{Spearman correlation (top) and Pearson correlation (bottom) between mean daily step counts from different algorithms and mean daily PA summaries. Correlations are not survey weighted since the interest is in correlation between raw estimates.}
    \label{fig:corr_means}
\end{figure}

Figure~\ref{fig:corr_means} displays the Spearman (top) and Pearson (bottom) correlations between mean daily step counts from each algorithm, MIMS, log10 MIMS, AC, and log10 AC. All Spearman correlations between step counting algorithms are larger than $0.8$, with the exception of Actilife and ADEPT ($0.77$).  The highest Spearman correlations are between Verisense and Verisense revised ($0.98$), Verisense revised and Oak ($0.97$), Verisense and Oak ($0.96$), and Actilife and Oak ($0.96$). Estimates of step counting algorithms are also highly correlated with widely used AC and MIMS measurements. The Spearman correlations are smaller for ADEPT (ranging from $0.5$ to $0.6$) and Stepcount SSL (ranging from $0.59$ to $0.7$). The highest Spearman correlations are between MIMS and Actilife ($0.93$), AC and Actilife ($0.92$), and Oak and MIMS ($0.90$). Interestingly, the Spearman correlation between AC and MIMS is extremely high ($0.99$) as is the correlation between log 10 AC and log 10 MIMS ($0.98$). The Pearson correlations reflect similar patterns and are most different from the Spearman correlations for ADEPT.

\subsection{Mortality prediction}\label{subsec:mortality_pred}
\subsubsection{Individual predictors of mortality}\label{subsubsec:univariate}
\begin{figure}[!htbp]
    \centering
    \includegraphics[width=0.9\textwidth]{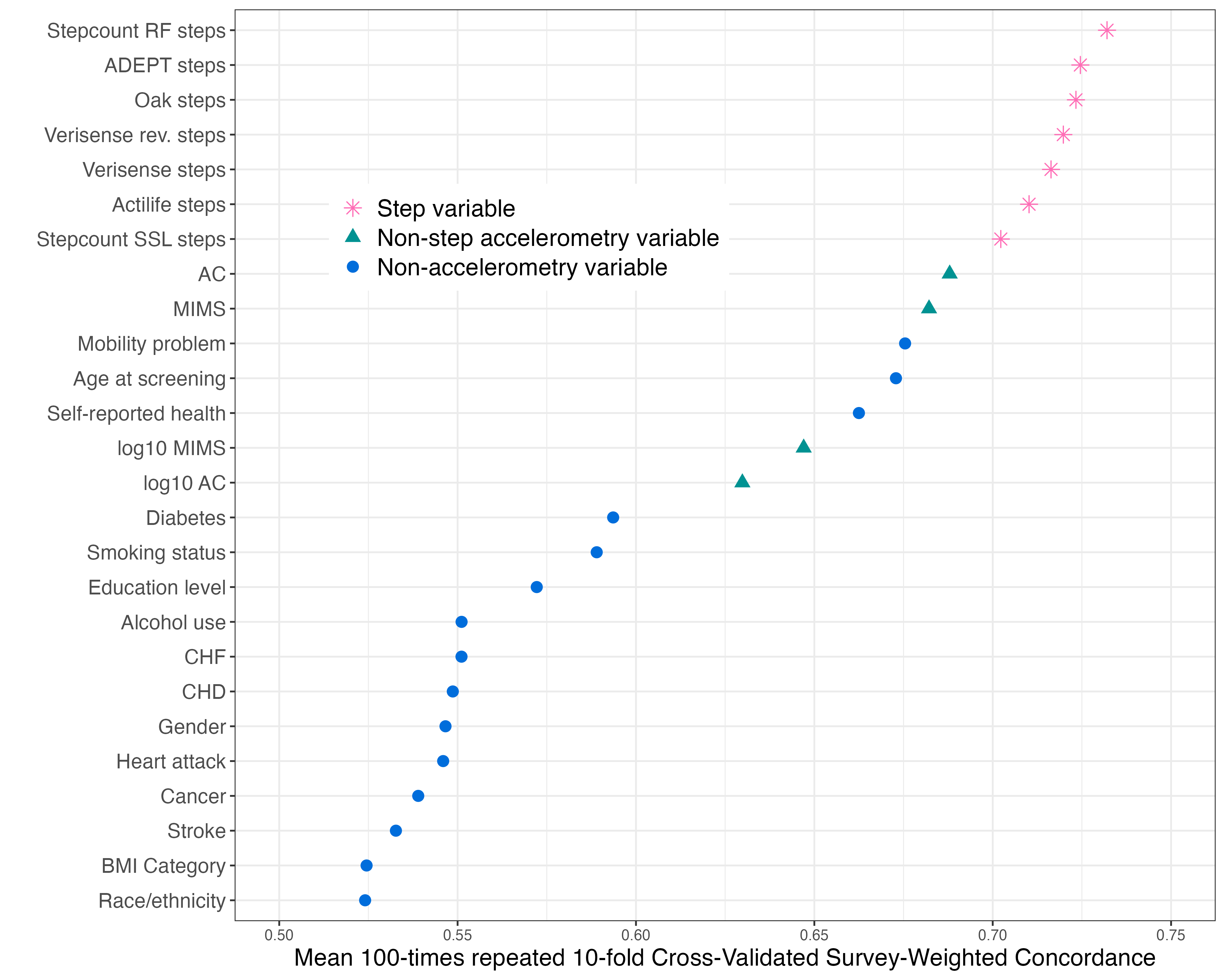}
    \caption{$100$ times repeated $10$-fold cross-validated survey weighted concordance from univariate Cox proportional hazards models.}
    \label{fig:single_concordance}
\end{figure}

Figure~\ref{fig:single_concordance} and Supplementary Table~\ref{tab:individual_concordance} present the $100$ times repeated $10$-fold cross validated survey weighted concordance from univariate Cox regression models. The mean or proportion of each variable by deceased and non-deceased groups are also presented. The nine variables with the highest concordance are all measures of physical activity; the seven models with the highest concordance are all step algorithms, followed by AC ($0.688$), MIMS ($0.682$), mobility problem ($0.675$), and age ($0.673$). The highest concordance corresponds to steps estimated with the Stepcount RF algorithm (C = $0.732$).

\subsubsection{The added predictive performance of PA summaries to the traditional risk factors of mortality}\label{subsubsec:comb}

For the multivariable mortality analysis, four models were fit. Each model included traditional predictors; the model with just these variables is referred to as ``traditional predictors." Then, three other models are considered: traditional predictors and Stepcount RF, traditional predictors and MIMS, and traditional predictors, Stepcount RF, and MIMS. Table~\ref{tab:multivar_summaries} presents the $100$ times repeated $10$-fold cross-validated survey weighted concordance for each model, along with the estimated coefficients and p-values from a model fit on all of the data. The concordance is highest for the model with traditional predictors and steps ($0.776$), though traditional predictors and MIMS ($0.773$) and traditional predictors, MIMS, and steps ($0.774$) were close. Furthermore, the coefficient for steps remains statistically significant even after the addition of MIMS, suggesting that total steps per day may confer additional information about mortality beyond that provided by MIMS and the traditional predictors used in the model. Furthermore, the confidence intervals for the hazard ratio associated with an 500-step increase overlap, indicating the effect of steps on mortality doesn't change significantly even after adjusting for MIMS.

Table~\ref{tab:adjusted_hr} presents the adjusted estimated hazard ratios and $95$\% confidence intervals associated with a $500$-step increase (raw) and a one standard deviation increase (scaled), which is algorithm-specific, in mean steps per day. The hazard ratios for all methods except ADEPT associated with a $500$-step increase in mean steps are between $0.95$ and $0.96$; ADEPT is $0.88$. Achieving a HR of 0.88 would be require a $11{,}000$ to $14{,}000$ increase in the number of steps for the other algorithms. The scaled hazard ratios associated with a one standard deviation increase are in close agreement for Actilife, ADEPT and Stepcount SSL ($0.67$), though a one standard deviation increase from ADEPT would correspond to an increase of $1{,}500$ steps, whereas all other algorithms would require an increase of at least $4{,}000$ steps to achieve the same reduction in risk.  Oak and Verisense revised also agree very closely with each other (hazard ratio $0.63$), though Oak would require an increase of almost $5{,}000$ steps whereas Verisense revised would require an increase of $4{,}000$ steps to achieve the same reduction in risk. Verisense estimates a hazard ratio of $0.65$ for an increase of one standard deviation (equivalent to $4{,}000$) steps, somewhere between Oak and Actilife. Stepcount RF estimates the smallest hazard ratio (most improvement) for an increase of one standard deviation (equivalent to $5{,}400$) steps. Note that all these algorithms are based on the same raw accelerometry data and differences are due only to the step counting algorithms. While ADEPT estimates a smaller number of steps than other algorithms for the same reduction in mortality risk, achieving an increase of $500$ ADEPT estimated steps may be as difficult as an increase of $2{,}000$ Oak estimated steps. Indeed, recall that we only have the outcomes of step counting algorithms and not the actual gold standard of the number of steps.

\section{Discussion}

We estimated steps in NHANES 2011-2014 using four open-source and one proprietary step counting algorithm based on the high-resolution three axial wrist accelerometry data. The open-source data (minute-level step counts for all algorithms and individuals in NHANES) published with this manuscript accompanied by a vignette detailing analyses in this paper will provide a valuable resource for future refinements of step counting algorithms and analysis of walking patterns among a nationally representative sample of Americans. 

There was substantial heterogeneity between step counting algorithms. The estimates for mean total steps among all adults by wave varied from $2{,}453$ (2013-14; ADEPT) to $12{,}169$ (2011-12; Actilife). However, the number of steps for all methods were very highly correlated with a minimum Spearman correlation of $0.77$, a median of $0.91$, and a maximum of $0.98$. In univariate Cox models predicting mortality, the mean total step count obtained by any algorithm outperformed all other accelerometry-derived summaries of PA (AC, MIMS, log10 AC, log10 MIMS) as well as traditional risk factors including age, self reported health, and mobility problem. 

In multivariable models, steps remained highly significant even after the inclusion of MIMS, the NHANES-provided summary of PA. For all step counting methods, an increase in steps was associated with a decrease in the hazard of mortality, but hazard ratios associated with a $500$-step increase varied from $0.88$ to $0.96$, again indicating the scale of the differences in these step estimates. The hazard ratios for a one standard deviation increase in the number of steps also varied between methods from $0.59$ to $0.67$. 

Analyses in this paper indicate that studies that use one of these step counting algorithms are likely to provide similar findings with respect to prediction of mortality. However, they also indicate that the estimators of the number of steps from different studies may not be comparable when using different algorithms or even slight modifications of the same algorithm. The most straightforward way to harmonize step counts across studies is to apply the same algorithm to the raw accelerometry data. From a translation perspective, choosing a particular step counting algorithm may have substantial effects on health recommendations and interventions. For example, if Stepcount RF were used to estimate the number of steps for health recommendations, the average number of steps for a person over the age of $50$ in the US would be around $12,000$ and an increase of $500$ steps per day would correspond to a $5$ percent reduction in risk (as measured by the mortality hazard). These estimators for the number of steps are larger than those published by Saint-Maurice \cite{saint-maurice_association_2020} (average of $9{,}124$ steps in US adults over the age of $40$), Tudor-Locke et al. \cite{tudor-locke_accelerometer-determined_2009} (average of $9{,}676$ steps in US adults over the age of $20$), and Small et al. \cite{stepcount} (average of $9{,}352$ in UK adults over the age of $40$). This number of steps could appear daunting for individuals over the age of $50$ and may result in discouraging people from even attempting to increase their PA. We argue that there is need to (1) closely investigating the accuracy of these estimates and identify potential sources of over-estimation of the number of steps; and (2) create a trusted source of open-source step counting algorithms that are version-controlled and calibrated to NHANES and, possibly, other studies.

Steps provide a measure of physical activity that is highly predictive of mortality. Indeed, our results support other findings on the relationship between step counts and mortality \cite{stepcount,lee_association_2019,master_association_2022,  hamer_dose-response_2022}.  MIMS and AC are widely used activity measures, but are not directly translatable to the public.  Minutes in conditions, such as light, moderate, and moderate-to-vigorous activity are translatable, but require thresholds and these thresholds have not been agreed upon. Steps do not require thresholding to estimate and are translatable to the general public. However, the large variation in estimates derived from different step counting algorithms limits exactly what is translated to the public, affects public health recommendations, and ultimately undermines the credibility of using step counts in research. Currently, we can recommend that ``more steps are better", but we cannot say how many more are optimal. More development is needed to validate existing step counting algorithms in free living settings; in order to do so, more free-living accelerometry datasets with ground truth step counts and activity labels are needed. 

Strengths of our analysis include the use of a large, nationally representative study. Our code is open-source and our analysis is reproducible. Limitations include the exclusion of individuals with insufficient wear; it is possible that these individuals have different patterns of PA, though assessing this exceeds the scope of this paper. Inaccuracy in the NHANES wear detection algorithm  could also have affected the inclusion of study participants. Furthermore, the association between steps and mortality found in our analysis should not be interpreted as causal and the short follow up time (median 6.75 years) in the study creates concern about reverse causality. Future directions include exploring the relationship between steps estimated by various step counting algorithms and mortality in other large, nationally representative studies. 

\section*{Acknowledgments}
This work was supported by the National Institutes of Health under Grant R01NS060910 and Grant R01AG075883. 
\section*{Conflict of interest}
Ciprian Crainiceanu is consulting for Bayer and Johnson and Johnson on
methods development for wearable and implantable technologies. The details of these contracts are disclosed through the Johns Hopkins University eDisclose system. The research presented here is not related to and was not supported by this consulting work.
\vspace{.1in}

The results of the study are presented clearly, honestly, and without fabrication, falsification, or inappropriate data manipulation. The results of the present study do not constitute endorsement by the American College of Sports Medicine.

\pagenumbering{gobble}
\section*{Figures} 

\clearpage 

\clearpage 

\section*{Tables} 

\begin{table}[!htbp]
\begin{tabular}{llll}
\textbf{Study} & \textbf{Target population} & \textbf{Method} & \textbf{Daily steps} \\
\hline\hline 
Saint Maurice\cite{saint-maurice_association_2020} & \begin{tabular}[c]{@{}l@{}}NHANES 2003-2006 (n=4840)\\ US adults aged 40+\end{tabular} & ActiGraph hip-worn & \begin{tabular}[c]{@{}l@{}}\textit{Mean} \\ 9124\end{tabular} \\
\hline 

Tudor Locke\cite{tudor-locke_accelerometer-determined_2009} & \begin{tabular}[c]{@{}l@{}}NHANES 2005-2006 (n=3744)\\ US adults aged 20+\end{tabular} & ActiGraph hip-worn & \begin{tabular}[c]{@{}l@{}}\textit{Mean (SE)} \\ 9676 (107) \\
6540 (106)$^*$ \end{tabular} \\
\hline 

Small\cite{stepcount} & \begin{tabular}[c]{@{}l@{}}UK Biobank (n=75943)\\ UK adults aged 40+\end{tabular} & \begin{tabular}[c]{@{}l@{}}Machine learning\\ (open-source)\end{tabular} & \begin{tabular}[c]{@{}l@{}}\textit{Median (IQR)} \\ 9352 (7099-11973)\end{tabular} \\
\hline 

Chan\cite{watchwalk} & \begin{tabular}[c]{@{}l@{}}UK Biobank (n=78822)\\ UK adults aged 40+\end{tabular} & \begin{tabular}[c]{@{}l@{}}Machine learning\\ (proprietary)\end{tabular} & \begin{tabular}[c]{@{}l@{}}\textit{Mean (SD)}\\ 8016 (3321)\end{tabular} \\
\hline 

Lee\cite{lee_association_2019} & \begin{tabular}[c]{@{}l@{}}Women's Health Study (n=16741)\\ US women aged 62+\end{tabular} & ActiGraph hip-worn & \begin{tabular}[c]{@{}l@{}}\textit{Mean} \\ 5499\end{tabular} \\
\hline 

Master\cite{master_association_2022} & \begin{tabular}[c]{@{}l@{}}All of Us (n=6042)\\ US adults median age 57\end{tabular} & Fitbit & \begin{tabular}[c]{@{}l@{}}\textit{Median (IQR)} \\ 7731 (5867-9827)\end{tabular}

\end{tabular}
\caption{Summary of literature estimating steps from large epidemiological studies\\
$*$: censored estimate; steps among minutes with $<$ 500 activity counts removed} 
\label{tab:methods}
\end{table}

\begin{table}[!htbp]
\centering
\renewcommand{\arraystretch}{.55}
\resizebox{\ifdim\width>\linewidth\linewidth\else\width\fi}{!}{
\begin{tabular}[t]{llll}
\toprule
& \textbf{Overall}& \textbf{2011-12} & \textbf{2013-14}\\
&  N = 8,664 &  N = 4,303 &  N = 4,361\\
\midrule
Female & 4,552 (53\%) & 2,281 (52\%) & 2,271 (53\%)\\
Age (yrs) & 48.01 (17.41) & 47.99 (17.29) & 48.04 (17.53)\\
Race/ethnicity &  &  & \\

\hspace{.1in}Non-Hispanic white & 5,785 (67\%) & 2,918 (67\%) & 2,867 (67\%)\\
\hspace{.1in}Non-Hispanic Black & 979 (11\%) & 505 (12\%) & 475 (11\%)\\
\hspace{.1in}Other or Multi-Race & 640 (7.4\%) & 312 (7.2\%) & 327 (7.6\%)\\
\hspace{.1in}Mexican American & 736 (8.5\%) & 344 (7.9\%) & 393 (9.1\%)\\
\hspace{.1in}Other Hispanic & 524 (6.0\%) & 288 (6.6\%) & 236 (5.5\%)\\

Education level &  &  & \\
\hspace{.1in}More than HS & 5,279 (63\%) & 2,664 (63\%) & 2,615 (63\%)\\
\hspace{.1in}HS/HS equivalent & 1,788 (21\%) & 877 (21\%) & 911 (22\%)\\
\hspace{.1in}Less than HS & 1,330 (16\%) & 703 (17\%) & 627 (15\%)\\
\hspace{.1in}\textit{Missing} & 267 & 123 & 144\\

BMI category &  &  & \\
\hspace{.1in}Normal & 2,436 (28\%) & 1,258 (29\%) & 1,178 (28\%)\\
\hspace{.1in}Obese & 3,176 (37\%) & 1,538 (36\%) & 1,639 (38\%)\\
\hspace{.1in}Overweight & 2,825 (33\%) & 1,445 (33\%) & 1,380 (32\%)\\
\hspace{.1in}Underweight & 156 (1.8\%) & 84 (2.0\%) & 71 (1.7\%)\\

\hspace{.1in}\textit{Missing} & 70 & 41 & 29\\
Diabetes & 887 (10\%) & 430 (9.9\%) & 457 (11\%)\\
\hspace{.1in}\textit{Missing} & 2 & 0 & 2\\
Coronary Heart Failure & 247 (2.9\%) & 135 (3.2\%) & 113 (2.7\%)\\
\hspace{.1in}\textit{Missing} & 270 & 127 & 143\\

Congenital Heart Disease & 316 (3.8\%) & 145 (3.4\%) & 171 (4.1\%)\\
\hspace{.1in}\textit{Missing} & 283 & 133 & 150\\
Stroke & 263 (3.1\%) & 140 (3.3\%) & 124 (3.0\%)\\
\hspace{.1in}\textit{Missing} & 267 & 124 & 143\\
Alcohol use &  &  & \\

\hspace{.1in}Never drinker & 1,057 (12\%) & 477 (11\%) & 581 (14\%)\\
\hspace{.1in}Former drinker & 1,233 (14\%) & 619 (14\%) & 614 (14\%)\\
\hspace{.1in}Moderate drinker & 2,657 (31\%) & 1,401 (32\%) & 1,256 (29\%)\\
\hspace{.1in}Heavy drinker & 669 (7.7\%) & 374 (8.6\%) & 296 (6.9\%)\\
\hspace{.1in}Missing alcohol & 3,048 (35\%) & 1,496 (34\%) & 1,552 (36\%)\\
Smoking status &  &  & \\
\hspace{.1in}Never smoker & 4,820 (56\%) & 2,351 (55\%) & 2,470 (57\%)\\
\hspace{.1in}Former smoker & 2,087 (24\%) & 1,071 (25\%) & 1,016 (24\%)\\
\hspace{.1in}Current smoker & 1,630 (19\%) & 820 (19\%) & 810 (19\%)\\
\hspace{.1in}\textit{Missing} & 127 & 126 & 1\\

Mobility problem & 1,411 (17\%) & 638 (15\%) & 773 (19\%)\\
\hspace{.1in}\textit{Missing} & 270 & 123 & 146\\
General health condition &  &  & \\
\hspace{.1in}Poor & 232 (2.7\%) & 106 (2.4\%) & 126 (2.9\%)\\
\hspace{.1in}Fair & 1,323 (15\%) & 612 (14\%) & 711 (17\%)\\

\hspace{.1in}Good & 3,408 (39\%) & 1,670 (38\%) & 1,738 (40\%)\\
\hspace{.1in}Very good & 2,742 (32\%) & 1,442 (33\%) & 1,300 (30\%)\\
\hspace{.1in}Excellent & 959 (11\%) & 536 (12\%) & 423 (9.8\%)\\
Died by 5 years follow up & 648 (7.5\%) & 354 (8.1\%) & 294 (6.8\%)\\
\hspace{.1in}\textit{Missing} & 15 & 12 & 4\\
\bottomrule
\end{tabular}}
\caption{Survey weighted population characteristics for individuals aged 18 and older with valid accelerometry data. Binary or categorical variables presented as n (\%) and continuous variables presented as mean (SD).}
\label{tab:pop_chars}
\end{table}

\begin{table}[!htbp]
\centering
\renewcommand{\arraystretch}{.8}
\resizebox{\ifdim\width>\linewidth\linewidth\else\width\fi}{!}{
\begin{tabular}{lllllll}
\toprule
\multicolumn{1}{c}{} & \multicolumn{2}{c}{2011-2012} & \multicolumn{2}{c}{2013-2014} & 
\multicolumn{2}{c}{\% Difference in Waves} \\
 & \textbf{Age 50+}, N = 2,107 & \textbf{All adults}, N = 4,303 & \textbf{Age 50+}, N = 2,146 & \textbf{All adults}, N = 4,361&
 \textbf{Age 50+} & \textbf{All adults}\\
\midrule
\textbf{Actilife} & 11,195 (4,001) & 12,169 (3,995) & 10,850 (4,008) & 11,902 (4,048) & 3.1 & 2.2 \\

\textbf{ADEPT} &2,342 (1,593) & 2,659 (1,569) & 2,151 (1,680) & 2,453 (1,575)  & 8.5 & 8.1 \\

\textbf{Oak} & 10,254 (5,027) & 11,794 (5,061) & 9,733 (4,886) & 11,381 (5,065)  & 5.2  & 3.6 \\

\textbf{Stepcount} & & & & \\
\hspace{.1in}RF & 9,888 (5,542) & 11,509 (5,722) & 9,502 (5,442) & 11,263 (5,764)  & 4.0 & 2.2 \\
\hspace{.1in}SSL &8,358 (4,402) & 9,144 (4,400) & 8,027 (4,388) & 8,846 (4,399)  & 4.0 & 3.3 \\

\textbf{Verisense} & & & & \\
\hspace{.1in} Original &8,337 (4,027) & 9,497 (4,062) & 7,974 (4,019) & 9,163 (4,065)  & 4.5 & 3.6 \\
\hspace{.1in} Revised & 7,532 (4,521) & 9,102 (4,756) & 7,122 (4,402) & 8,725 (4,730)  & 5.6 & 4.2 \\
\hline 
\textbf{MIMS} & & & & \\
\hspace{.1in}Raw & 12,435 (3,642) & 13,572 (3,758) & 12,243 (3,574) & 13,467 (3,784) & 1.6 & 0.78 \\
\hspace{.1in}log10 &  960 (158) & 998 (154) & 952 (160) & 994 (155)  & 0.84 & 0.40 \\
\textbf{AC} & & & & \\
\hspace{.1in}Raw & 2,333,927 (791,711) & 2,573,262 (815,192) & 2,291,942 (766,303) & 2,549,846 (816,305) & 1.8 & 0.91 \\
\hspace{.1in}log10 &2,878 (383) & 2,955 (366) & 2,847 (394) & 2,933 (370)  & 1.1 & 0.75 \\
\bottomrule
\end{tabular}}
\caption{Survey weighted mean (SD) physical activity totals by wave and age among individuals with at least three valid days of accelerometry data. The final column shows the percent difference in estimates between 2011-2012 and 2013-2014: $\frac{|est_{2011-12} - est_{2013-14}|}{mean(est_{2011-12}, est_{2013-14})}$}
\label{tab:steps_pa}
\end{table}

\clearpage 

\begin{table}[ht]
\centering 
\renewcommand{\arraystretch}{.9}
\resizebox{\ifdim\width>\linewidth\linewidth\else\width\fi}{!}{
\begin{tabular}[t]{llll}
\hline
Model & Steps Hazard Ratio$^*$ & Steps p-value & Model Concordance\\
\hline \hline
Traditional predictors only & --- & ---& 0.769 \\

Traditional predictors + MIMS & --- & --- & 0.773\\
Traditional predictors + Stepcount RF steps & 0.955 (0.940, 0.970) & $<$0.001 & 0.776 \\
Traditional predictors + Stepcount RF steps + MIMS & 0.961 (0.939, 0.983) & 0.0361 & 0.774 \\
\hline
\end{tabular}}
\caption{Multivariable model summaries. For each model, the hazard ratio and associated p-value for steps per day is reported, along with the average of $100$ times repeated 10-fold cross-validated model concordance. \\
$^*$Hazard ratio associated with an increase of $500$ steps per day}
\label{tab:multivar_summaries}
\end{table}

\newpage 

\begin{table}[ht]
\centering
\renewcommand{\arraystretch}{1}
\resizebox{\ifdim\width>\linewidth\linewidth\else\width\fi}{!}{
\begin{tabular}[t]{llll}
\toprule
Step algorithm& Adjusted HR; raw &Adjusted HR; scaled & SD of Steps (x1000)\\
\midrule
Actilife &  0.95 (0.93, 0.97) & 0.67 (0.58, 0.78) & 4.0\\
ADEPT & 0.88 (0.83, 0.93) & 0.67 (0.56, 0.80) & 1.5\\
Oak &  0.95 (0.94, 0.97) & 0.63 (0.54, 0.74) & 4.9\\
Verisense & 0.95 (0.93, 0.97) & 0.65 (0.55, 0.76) & 4.0\\
Verisense rev. &  0.95 (0.93, 0.97) & 0.63 (0.54, 0.75) & 4.4\\
Stepcount RF &  0.95 (0.94, 0.97) & 0.59 (0.50, 0.70) & 5.4\\
Stepcount SSL &  0.96 (0.94, 0.97) & 0.67 (0.57, 0.79) & 4.4\\
\bottomrule
\end{tabular}}

\caption{Adjusted hazard ratios and associated 95\% confidence intervals associated with a 500-unit increase in steps (raw) and one standard deviation increase in steps (scaled). The standard deviation is in thousands of steps and can be considered what a one unit increase in the scaled predictor predictor represents in terms of steps. Hazard ratios are obtained from weighted Cox proportional hazards regression models adjusting for the traditional predictors described: age, sex, BMI, race, diabetes, CHF, self-reported health, CHD, heart attack, stroke, cancer, alcohol use, smoking, mobility problem, and education.} 
\label{tab:adjusted_hr}
\end{table}

\clearpage

\appendix

\renewcommand\thesection{}  

\renewcommand\thefigure{S\arabic{figure}}  
\renewcommand\thetable{S\arabic{table}}    

\setcounter{figure}{0}
\setcounter{table}{0}

\section{Supplemental Figures and Tables}

\begin{figure}[!htbp]
    \centering
    \includegraphics[width=0.95\linewidth]{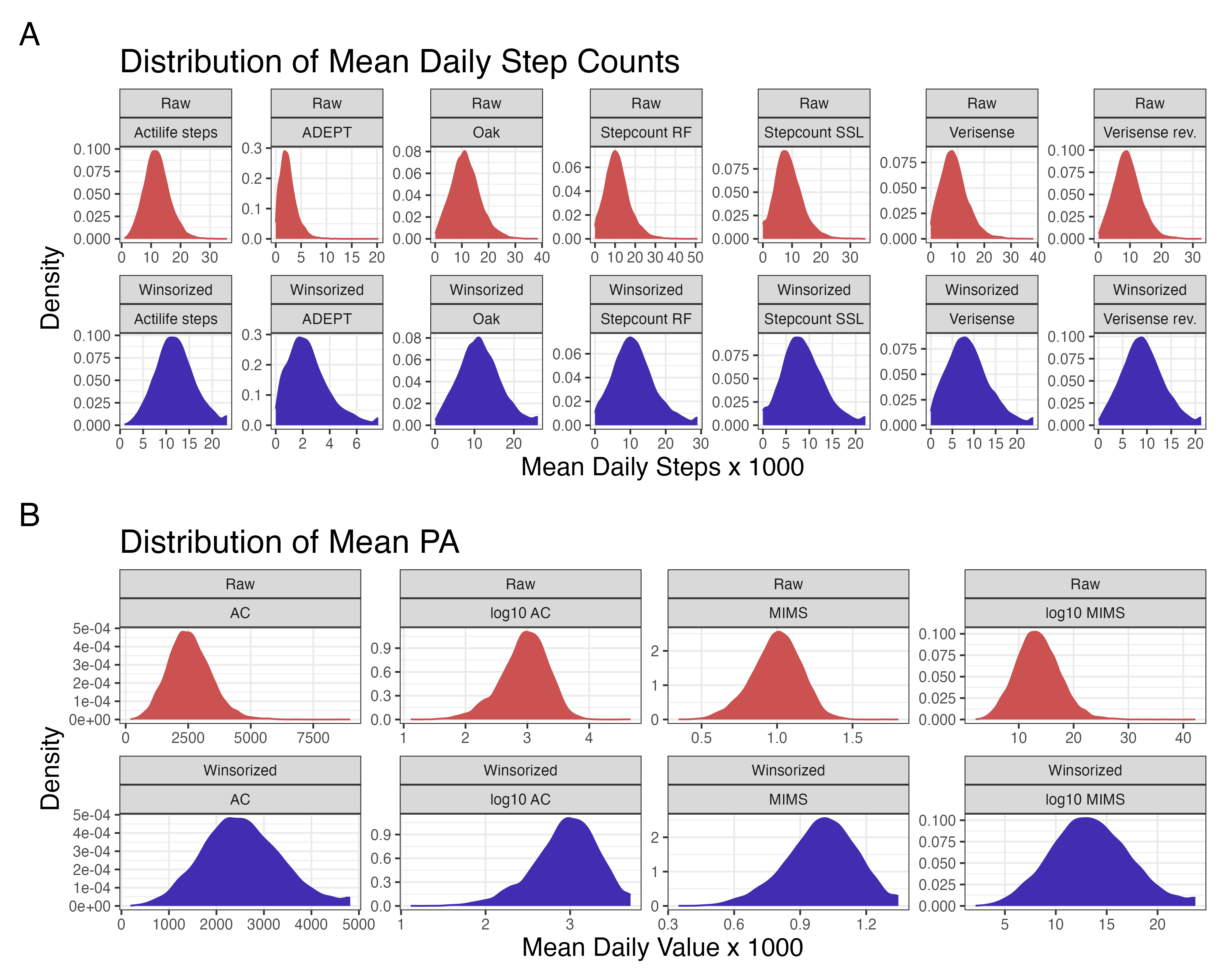}
    \caption{Panel \textbf{A}: Raw (top) and Winsorized (bottom) distributions for steps among all individuals 18 and older with valid accelerometry data. Panel \textbf{B}: raw (top) and Winsorized (bottom) and AC, log10 AC and MIMS, and log10 MIMS among all individuals 18 and older with valid accelerometry data}
   \label{fig:distributions}
\end{figure}
\clearpage

\begin{figure}[ht]
    \centering
    \includegraphics[width=\linewidth]{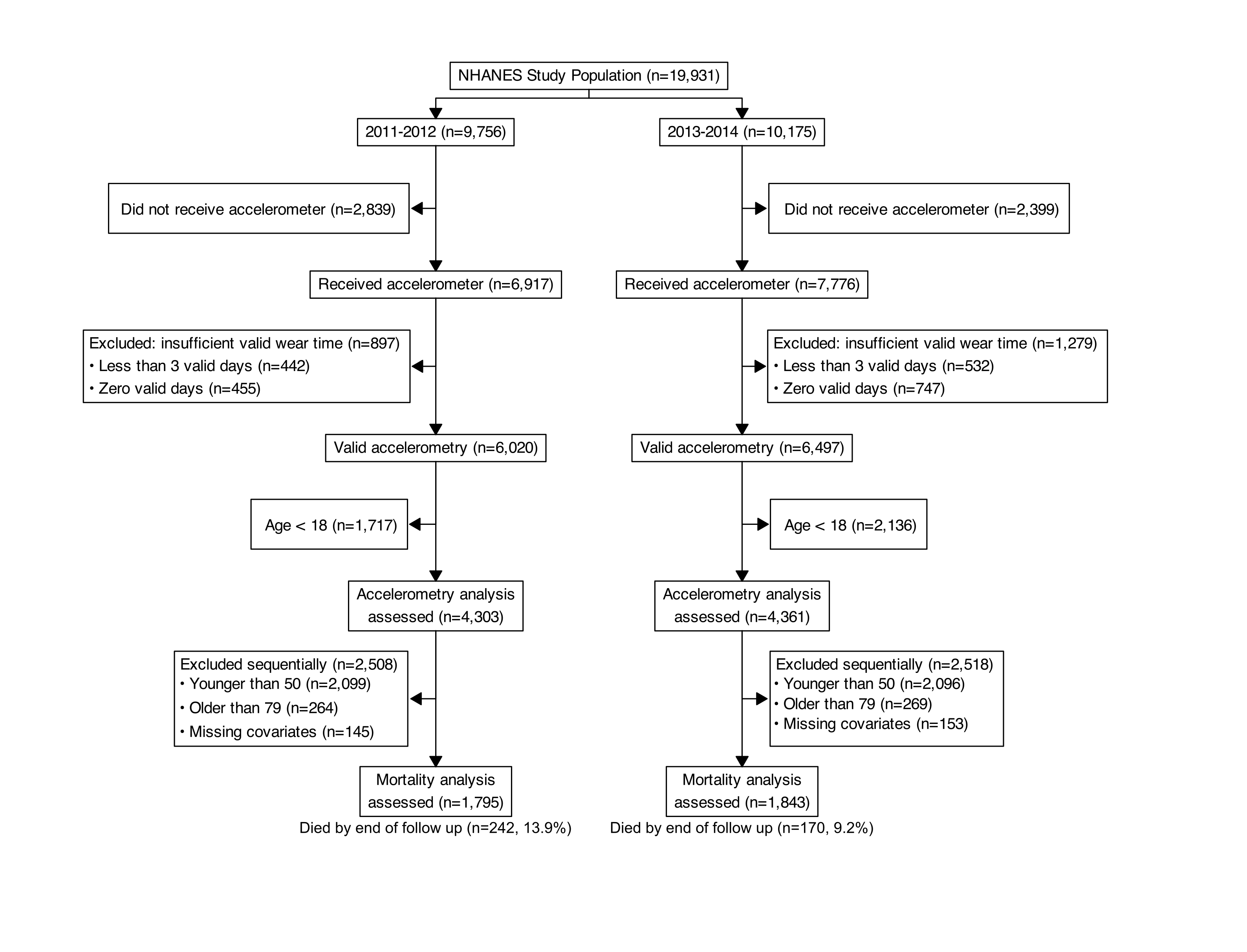}
    \caption{Inclusion Diagram}
    \label{fig:supp-consort}
\end{figure}

\clearpage 
\begin{table}[ht]
\centering
\renewcommand{\arraystretch}{1}
\resizebox{\ifdim\width>\linewidth\linewidth\else\width\fi}{!}{
\begin{tabular}[t]{lll}
\toprule
Step algorithm& Time for 10 subjects (minutes) &Estimated time for entire data (days) \\
\midrule
ADEPT* & 716 &  731 \\
Oak & 69.7 &  71 \\
Stepcount RF & 53.8 &  54 \\
Stepcount SSL & 52.7 & 53 \\
Verisense & 18.0 &  18\\
Verisense rev. & 46.9& 48 \\
\midrule 
\textbf{All algorithms} & 902 & 920 \\
\bottomrule
\end{tabular}}
\caption{Computation time required to run step counting algorithms on a standard computer. The time for 10 subjects is calculated by fitting the models on a random sample of ten subjects. The time to run on all of NHANES is estimated by multiplying the time for 10 subjects by $6917+7776$ (number of subjects who received accelerometers in NHANES) and dividing by $60*24 = 1440$ to convert from minutes to days. \\
*ADEPT run in parallel using 8 cores}
\label{tab:compute}
\end{table}
\clearpage 

\begin{table}[ht]
\begin{tabular}{@{}clllll@{}}
\toprule
\multicolumn{1}{l}{} &  & \multicolumn{4}{c}{Preceding state} \\ \cmidrule(l){2-6} 
\multicolumn{1}{l|}{} &  & Unknown & Nonwear & Sleep wear & Wake wear \\
\multicolumn{1}{c|}{\multirow{4}{*}{Following state}} & Unknown & -- & 0.005 & 0.006 & 0.005 \\
\multicolumn{1}{c|}{} & Nonwear & 0.012 & 0.021 & 0.005 & 0.011 \\
\multicolumn{1}{c|}{} & Sleep wear & 0.004 & 0.006 & 0.280 & 0.123 \\
\multicolumn{1}{c|}{} & Wake wear & 0.004 & 0.013 & 0.12 & 0.457
\end{tabular}
\caption{Probability matrix for minute-level wear predictions preceding and following a bout of ``unknown" wear predictions. The majority of time, unknown falls between two bouts of wake wear (46\%), two bouts of sleep wear (28\%), or a bout of sleep wear and wake wear (24\%).}
    \label{tab:tpm}
\end{table}

\begin{table}[ht]
\centering
\renewcommand{\arraystretch}{.5}
\resizebox{\ifdim\width>\linewidth\linewidth\else\width\fi}{!}{
\begin{tabular}[t]{lccc}
\toprule

\multirow{3}{*}{Variable} & \multicolumn{3}{c}{Concordance} \\
\cmidrule{2-4}
& 50-79 years & 50-79 years& $50+$ years\\
&$\geq 3$ valid days & $\geq 1$ valid days & $\geq 3$ valid days\\
\midrule
Stepcount RF steps & 0.732 & 0.732 & 0.776\\
ADEPT steps & 0.725 & 0.724 & 0.768\\
Oak steps & 0.723 & 0.723 & 0.767\\
Verisense rev. steps & 0.720 & 0.719 & 0.763\\
Verisense steps & 0.716 & 0.716 & 0.758\\

Actilife steps & 0.710 & 0.711 & 0.748\\
Stepcount SSL steps & 0.702 & 0.703 & 0.747\\
AC & 0.688 & 0.689 & 0.728\\
MIMS & 0.682 & 0.683 & 0.722\\
Mobility problem & 0.675 & 0.675 & 0.686\\

Age (yrs) & 0.673 & 0.672 & 0.751\\
General health condition & 0.662 & 0.664 & 0.633\\
log10 MIMS & 0.647 & 0.650 & 0.689\\
log10 AC & 0.630 & 0.633 & 0.672\\
Diabetes & 0.594 & 0.591 & 0.562\\

Smoking status & 0.589 & 0.587 & 0.553\\
Education category & 0.572 & 0.574 & 0.575\\
Alcohol use & 0.551 & 0.550 & 0.594\\
Congenital Heart Disease & 0.551 & 0.551 & 0.557\\
Coronary Heart Failure & 0.549 & 0.549 & 0.560\\

Sex & 0.547 & 0.546 & 0.523\\
Heart attack & 0.546 & 0.546 & 0.546\\
Cancer & 0.539 & 0.538 & 0.549\\
Stroke & 0.533 & 0.532 & 0.539\\
BMI category & 0.524 & 0.519 & 0.530\\

Race/ethnicity & 0.524 & 0.526 & 0.518\\
\bottomrule
\end{tabular}}
\caption{Sensitivity analysis comparing 100-times repeated 10-fold cross validated concordance for single variable models including individuals with $\geq 3$ valid days of data between 50 and 79 years old, individuals with $\geq 1$ valid day of data between 50 and 79 years old, and individuals with $\geq 3$ valid days of data age 50 and older, respectively.}
\label{tab:sens_univariate}
\end{table}

\clearpage

\begin{table}[ht]
\centering
\renewcommand{\arraystretch}{.5}
\resizebox{\ifdim\width>\linewidth\linewidth\else\width\fi}{!}{
\begin{tabular}[t]{lcccccc}
\toprule 
\multirow{3}{*}{Variable} & \multicolumn{3}{c}{Concordance} & \multicolumn{3}{c}{Hazard Ratio} \\
\cmidrule(lr){2-4} \cmidrule(lr){5-7} 
& 50-79 years  & 50-79 years & $50+$ years & 50-79 years  & 50-79 years & $50+$ years\\
& $\geq 3$ valid days &  $\geq 1$ valid days & $\geq 3$ valid days & $\geq 3$ valid days & $\geq 1$ valid days & $\geq 3$ valid days\\
\midrule
Oak  & 0.777 & 0.777 & 0.813 & 0.955 & 0.953 & 0.944\\
Actilife  & 0.776 & 0.776 & 0.812 & 0.952 & 0.950 & 0.947\\
Stepcount RF  & 0.776 & 0.776 & 0.814 & 0.952 & 0.950 & 0.941\\
Verisense rev.  & 0.776 & 0.775 & 0.813 & 0.950 & 0.949 & 0.938\\
Verisense & 0.775 & 0.775 & 0.812 & 0.947 & 0.945 & 0.937\\
AC & 0.774 & 0.774 & 0.811 & --  & --  & -- \\
MIMS & 0.773 & 0.773 & 0.810 & --  & --  & -- \\
Stepcount SSL & 0.773 & 0.773 & 0.811 & 0.956 & 0.953 & 0.942\\
ADEPT & 0.772 & 0.772 & 0.809 & 0.879 & 0.874 & 0.846\\
log10 MIMS& 0.771 & 0.770 & 0.807 & --  & --  & -- \\

log10 AC & 0.770 & 0.770 & 0.807 & --  & --  & -- \\
Traditional predictors only & 0.769 & 0.768 & 0.805 & --  & --  & -- \\
\bottomrule
\end{tabular}}
\caption{Sensitivity analysis comparing 100-times repeated 10-fold cross validated concordance and estimated hazard ratio of mortality associated with a 500-step increase for multivariable models including individuals with $\geq 3$ valid days of data between 50 and 79 years old, individuals with $\geq 1$ valid day of data between 50 and 79 years old, and individuals with $\geq 3$ valid days of data age 50 and older, respectively. All models control for traditional predictors in addition to the physical activity variable. All models control for traditional predictors and the given physical activity variable.}
\label{tab:sens_multivar}
\end{table}

\clearpage
\begin{table}[ht]
\centering
\renewcommand{\arraystretch}{.5}
\resizebox{\ifdim\width>\linewidth\linewidth\else\width\fi}{!}{
\begin{tabular}[t]{llll}
\toprule
& \textbf{Overall}& \textbf{2011-12} & \textbf{2013-14}\\
&  N = 3,720 & N = 1,843  &  N = 1,877\\
\midrule
Female & 1,948 (52\%) & 958 (52\%) & 990 (53\%)\\
Age (yrs) & 62.70 (8.04) & 62.62 (8.12) & 62.79 (7.97)\\
Race/ethnicity &  &  & \\

\hspace{.1in}Non-Hispanic white & 1,487 (40\%) & 660 (36\%) & 827 (44\%)\\
\hspace{.1in}Non-Hispanic Black & 999 (27\%) & 567 (31\%) & 432 (23\%)\\
\hspace{.1in}Other or Multi-Race & 451 (12\%) & 244 (13\%) & 207 (11\%)\\
\hspace{.1in}Mexican American & 395 (11\%) & 149 (8.1\%) & 246 (13\%)\\
\hspace{.1in}Other Hispanic & 388 (10\%) & 223 (12\%) & 165 (8.8\%)\\

Education level &  &  & \\
\hspace{.1in}More than HS & 1,894 (51\%) & 929 (50\%) & 965 (51\%)\\
\hspace{.1in}Less than HS & 979 (26\%) & 507 (28\%) & 472 (25\%)\\
\hspace{.1in}HS/HS equivalent & 845 (23\%) & 407 (22\%) & 438 (23\%)\\
\hspace{.1in}\textit{Missing} & 2 & 0 & 2\\

BMI category &  &  & \\
\hspace{.1in}Normal & 904 (25\%) & 463 (25\%) & 441 (24\%)\\
\hspace{.1in}Obese & 1,519 (41\%) & 734 (40\%) & 785 (42\%)\\
\hspace{.1in}Overweight & 1,207 (33\%) & 599 (33\%) & 608 (33\%)\\
\hspace{.1in}Underweight & 57 (1.5\%) & 22 (1.2\%) & 35 (1.9\%)\\

\hspace{.1in}\textit{Missing} & 33 & 25 & 8\\
Diabetes & 828 (22\%) & 414 (22\%) & 414 (22\%)\\
\hspace{.1in}\textit{Missing} & 3 & 1 & 2\\
Coronary Heart Failure & 183 (4.9\%) & 97 (5.3\%) & 86 (4.6\%)\\
\hspace{.1in}\textit{Missing}& 13 & 8 & 5\\

Congenital Heart Disease & 231 (6.2\%) & 106 (5.8\%) & 125 (6.7\%)\\
\hspace{.1in}\textit{Missing} & 17 & 9 & 8\\
Stroke & 222 (6.0\%) & 125 (6.8\%) & 97 (5.2\%)\\
\hspace{.1in}\textit{Missing} & 4 & 0 & 4\\
Alcohol use &  &  & \\

\hspace{.1in}Never drinker & 551 (15\%) & 279 (15\%) & 272 (14\%)\\
\hspace{.1in}Former drinker & 863 (23\%) & 419 (23\%) & 444 (24\%)\\
\hspace{.1in}Moderate drinker & 847 (23\%) & 413 (22\%) & 434 (23\%)\\
\hspace{.1in}Heavy drinker & 238 (6.4\%) & 128 (6.9\%) & 110 (5.9\%)\\
\hspace{.1in}Missing alcohol & 1,221 (33\%) & 604 (33\%) & 617 (33\%)\\

Smoking status &  &  & \\
\hspace{.1in}Never smoker & 1,831 (49\%) & 922 (50\%) & 909 (48\%)\\
\hspace{.1in}Former smoker & 1,204 (32\%) & 592 (32\%) & 612 (33\%)\\
\hspace{.1in}Current smoker & 683 (18\%) & 328 (18\%) & 355 (19\%)\\
\hspace{.1in}\textit{Missing} & 2 & 1 & 1\\

Mobility problem & 1,126 (30\%) & 534 (29\%) & 592 (32\%)\\
\hspace{.1in}\textit{Missing}& 4 & 0 & 4\\
General health condition &  &  & \\
\hspace{.1in}Poor & 213 (5.7\%) & 103 (5.6\%) & 110 (5.9\%)\\
\hspace{.1in}Fair & 908 (24\%) & 444 (24\%) & 464 (25\%)\\

\hspace{.1in}Good & 1,457 (39\%) & 711 (39\%) & 746 (40\%)\\
\hspace{.1in}Very good & 870 (23\%) & 437 (24\%) & 433 (23\%)\\
\hspace{.1in}Excellent & 272 (7.3\%) & 148 (8.0\%) & 124 (6.6\%)\\
Died by 5 years follow up & 440 (12\%) & 261 (14\%) & 179 (9.6\%)\\
\hspace{.1in}\textit{Missing} & 9 & 5 & 4\\
\bottomrule
\end{tabular}}
\caption{Unweighted population characteristics for individuals aged 18-79 who had valid accelerometry data.}
\label{tab:mortality_tab1}
\end{table}

\clearpage

\begin{table}[ht]
\centering
\renewcommand{\arraystretch}{.45}
\resizebox{\ifdim\width>\linewidth\linewidth\else\width\fi}{!}{
\begin{tabular}[t]{lccc}
\toprule

\multirow{2}{*}{Variable} & \multirow{2}{*}{Concordance} & \multicolumn{2}{c}{Mean or \%} \\
\cmidrule{3-4} 
& & Alive & Deceased\\
\midrule

Stepcount RF steps & 0.732 & 10,591 & 6,502 \\
ADEPT steps & 0.725& 2,391 & 1,380 \\
Oak steps & 0.723 & 10,845 & 7,067 \\
Verisense rev. steps & 0.720  & 8,021 & 4,878 \\
Verisense steps & 0.716  & 8,885 & 5,885 \\

Actilife steps & 0.710 & 11,790 & 8,740 \\
Stepcount SSL steps & 0.702  & 8,786 & 5,784 \\
AC & 0.688  & 2,454,243 & 1,900,197 \\
MIMS & 0.682 & 13,002 & 10,438 \\
Mobility problem (\%) & 0.675  & 26 & 57 \\

Age (yrs) & 0.673  & 62 & 68 \\
General health condition (\%) & 0.662&  &  \\
\hspace{.1in}Poor & & 4.4 & 13 \\
\hspace{.1in}Fair & & 23 & 36 \\
\hspace{.1in}Good & & 40 & 33 \\
\hspace{.1in}Very good & & 25 & 13 \\
\hspace{.1in}Excellent & & 7.8 & 4.6 \\
log10 MIMS & 0.647  & 987 & 884 \\
log10 AC & 0.630  & 2,939 & 2,708 \\
Diabetes (\%)& 0.594  & 20 & 37 \\

Smoking status (\%)& 0.589  &  &  \\
\hspace{.1in}Never & & 51 & 36 \\
\hspace{.1in}Former & & 31 & 40  \\
\hspace{.1in}Current & & 18 & 24  \\
Education category (\%) & 0.572  &  &  \\
\hspace{.1in}More than HS & & 53 & 41 \\
\hspace{.1in}HS/HS equivalent& & 23 & 24 \\
\hspace{.1in}Less than HS & & 25 & 35 \\

Alcohol use (\%)& 0.551  &  &  \\
\hspace{.1in}Never drinker & & 15 & 11 \\
\hspace{.1in}Former drinker & & 21 & 36 \\
\hspace{.1in}Moderate drinker & & 23 & 18 \\
\hspace{.1in}Heavy drinker & & 6.3 & 7.5 \\
\hspace{.1in}Missing alcohol & & 33 & 27 \\
Congenital Heart Disease (\%)& 0.551  & 5.0 & 15 \\
Coronary Heart Failure (\%)& 0.549  & 3.4 & 15\\

Female (\%)& 0.547  & 53 & 44 \\
Heart attack (\%)& 0.546 & 5.1 & 16 \\
Cancer (\%)& 0.539  & 14 & 24 \\
Stroke (\%)& 0.533 & 4.7 & 13 \\
BMI category (\%)& 0.524  &  &  \\
\hspace{.1in}Normal & & 24 & 26 \\
\hspace{.1in}Obese & & 41 & 42 \\
\hspace{.1in}Overweight & & 36 & 28 \\
\hspace{.1in}Underweight & & 1.3 & 3.4 \\
Race/ethnicity (\%)& 0.524 &  &  \\
\hspace{.1in}Non-Hispanic white & & 39 & 47 \\
\hspace{.1in}Non-Hispanic Black & & 26 & 30 \\
\hspace{.1in}Other or Multi-Race & & 13 & 8.3 \\
\hspace{.1in}Mexican-American & & 11 & 8.0 \\
\hspace{.1in}Other Hispanic & & 11 & 6.8 \\
\bottomrule
\end{tabular}}
\caption{100 times repeated 10-fold cross validated survey weighted concordance for individual predictors of mortality in Cox proportional hazards models. The mean of continuous variables and percent for binary variables for individuals who were alive and who were deceased at the end of follow up, respectively, are also displayed.}
\label{tab:individual_concordance}
\end{table}

\clearpage

\clearpage

\bibliographystyle{unsrt}

\bibliography{references}
\end{document}